\documentclass[aps,prl,twocolumn,showpacs]{revtex4}
\usepackage{graphicx}
\usepackage{hyperref}

\begin{document}

\title{Weakly interacting two-dimensional system of dipoles:
limitations of mean-field theory}

\author{G.E.~Astrakharchik}
\affiliation{Departament de F\'{\i}sica i Enginyeria Nuclear, Campus Nord B4-B5, Universitat Polit\`ecnica de Catalunya, E-08034 Barcelona, Spain}

\author{J.~Boronat}
\affiliation{Departament de F\'{\i}sica i Enginyeria Nuclear, Campus Nord B4-B5, Universitat Polit\`ecnica de Catalunya, E-08034 Barcelona, Spain}

\author{J.~Casulleras}
\affiliation{Departament de F\'{\i}sica i Enginyeria Nuclear, Campus Nord B4-B5, Universitat Polit\`ecnica de Catalunya, E-08034 Barcelona, Spain}

\author{I.L.~Kurbakov}
\affiliation{Institute of Spectroscopy, 142190 Troitsk, Moscow region, Russia}

\author{Yu.E.~Lozovik}
\affiliation{Institute of Spectroscopy, 142190 Troitsk, Moscow region, Russia}

\date{\today}

\begin{abstract}
We consider a homogeneous 2D Bose gas with repulsive dipole-dipole interactions. The
ground-state equation of state, calculated using the Diffusion Monte Carlo method,
shows quantitative differences with predictions of commonly used Gross-Pitaevskii
mean-field theory. The static structure factor, pair distribution function and
condensate fraction are calculated in a wide range of the gas parameter. Differences
with mean-field theory are reflected in the frequency of the lowest ``breathing''
mode for harmonically trapped systems.
\end{abstract}


\maketitle

The study of quasi-two-dimensional Bose gases at ultra-low temperatures has become a
very active area of research. The role of correlations and quantum fluctuations is
greatly enhanced in reduced dimensionality making a two-dimensional (2D) system well
suited for studying beyond mean-field effects. The superfluid-normal phase
transition occurs at a finite-temperature and follows the peculiar scenario of
Berezinskii, Kosterlitz, Thouless \cite{BKT}. On the contrary, the system undergoes
Bose-Einstein condensation (BEC) only at zero temperature\cite{Hohenberg67}, since
long-wavelength phase fluctuations destroy long-range order.
The analytical descriptions of 2D systems include mean-field Gross-Pitaevskii (GPE)
theory \cite{Schick71,GPE}, beyond mean-field approaches \cite{BMF,Posazhennikova06},
and numerical methods \cite{DipNum,Astrakharchik06,buchler06}.

Recently, Bose-Einstein condensation was achieved in $^{52}$Cr atoms
\cite{BECatoms}. Chromium possesses a large permanent magnetic moment making the
experimental observation of effects of dipole-dipole interactions
possible \cite{Stuhler05}. Moreover, by tuning its $s$-wave scattering length by a
Fesh\-bach resonance, one can realize a system with purely dipolar interactions. The
first study of the Feshbach resonances in $^{52}$Cr atoms appeared recently
\cite{Werner05}. As shown in Ref.~\cite{Petrov00b}, this procedure would permit to
tune the effective quasi-two-dimensional coupling constant $g_{2D}$. In addition,
the Bose-Einstein condensation of 2D dipolar excitons in quantum wells was recently
observed in luminescence experiments \cite{exc}. In two coupled quantum wells, one
containing only holes, and the other only electrons, holes and electrons might
couple forming indirect excitons. Alternatively, the exciton dipole moment can be
induced by normal electric field in a single quantum well. Spatial separation
between hole and electron suppresses recombination and greatly increases the
lifetime of an exciton. If the separation between excitons are greater than the
electron - hole separation $D$, an indirect exciton can be approximated as a boson
with dipolar moment oriented perpendicularly to the plane.

A homogeneous system of $N$ dipoles is described by the Hamiltonian
\begin{eqnarray}
\hat H = -\frac{\hbar^2}{2M}\sum_{i=1}^N \Delta_i +
\frac{C_{dd}}{4\pi}\sum_{j<k}\frac{1}{|{\bf r_j - r_k}|^3},
\label{H}
\end{eqnarray}
where $M$ is the mass of a dipole. The coupling constant $C_{dd}$ depends on the
nature of the dipolar interaction. For cold atoms with large permanent magnetic
moment $m$ one has $C_{dd}=\mu_0m^2$, where $\mu_0$ is the permeability of free
space. If the dipole moment is induced by an electric field $E$, the coupling
constant is $C_{dd} = E^2\alpha^2/\epsilon_0$, where $\alpha$ is the static
polarizability and $\epsilon_0$ the permittivity of free space. In the case of
excitons $C_{dd} = e^2D^2/\varepsilon$, where $e$ is the electron charge and
$\varepsilon$ is the dielectric constant of the semiconductor.

We suppose that the atoms are confined in a very tight pancake trap, so that
dynamically the system is two-dimensional, and the dipole moments are aligned
perpendicularly to the plane. This stabilizes the system since interactions are
purely repulsive.

The properties of Hamiltonian (\ref{H}) are governed by the two-dimensional gas
parameter $na^2$, $n$ being the 2D density and $a$ the 2D $s$-wave scattering
length. The limit $na^2\to 0$ corresponds to a weakly interacting regime where the
mean-field (MF) theory can be applied. The MF energy per particle is proportional to
the coupling constant: $E^{MF}/N = g n/2$. The leading contribution to the purely 2D
coupling constant was first obtained by Shick\cite{Schick71} and rigorously derived
for GPE by Lieb {\it et al.}\cite{GPE}
\begin{eqnarray}
g^{MF}_{2D} = \frac{4\pi\hbar^2}{M}\frac{1}{|\ln na^2|}
\label{Emf}
\end{eqnarray}
A peculiarity of a two-dimensional system is that the coupling constant depends on
density, while it is independent of density in three- and one- dimensional systems.
This factor causes particular energetic properties (for example, in the frequencies
of collective oscillations). Calculation of subleading corrections to MF (\ref{Emf})
is a difficult and long-standing problem and it was addressed by different authors
\cite{BMF} giving sometimes controversial results.

In a 2D system a Bose-Einstein condensate can only be formed at zero temperature.
The system is completely condensed in the limit $na^2\to 0$, while in a denser
system the condensate density $n_0$ gets depleted. The departure from the fully
condensed state is given by the perturbative expression \cite{Schick71}
\begin{eqnarray}
\frac{n_0}{n} = 1 - \frac{1}{|\ln na^2 |}+{\cal O}\left(\frac{1}{\ln^2 na^2}\right)
\label{CF}
\end{eqnarray}

For small momenta the excitation spectrum is linear ${\cal E}_k=\hbar |k| c$, with $c$
being the speed of sound, which corresponds to having phonons in the system. This
defines a linear behavior of the static structure factor
\begin{eqnarray}
S^{ph.}_k=\frac{\hbar |k|}{2Mc},\qquad |k|/n^{1/2}\ll 1
\label{Skph}
\end{eqnarray}
in agreement with Feynman formula \cite{Feynman54} that relates the excitation
spectrum to the static structure factor ${\cal E}_k = \hbar^2 k^2 / (2M S_k)$.
Excitations at very high momenta $|k|/n^{1/2}\gg 1$ are free particles ${\cal E}_k =
\hbar^2 k^2 / 2M$ and $S_k = 1$. The intermediate behavior of the static structure
factor at $|k|n^{1/2}\approx 1$ can be approximated in the weakly-interacting regime
by assuming the Bogoliubov excitation spectrum
${\cal E}_k=\sqrt{\hbar^2k^2\mu/M+(\hbar^2k^2/2M)^2}$,
\begin{eqnarray}
S^{Bog.}_k = \frac{|k|}{\sqrt{16\pi n/|\ln na^2|+k^2}},\qquad na^2\ll 1
\label{SkBog}
\end{eqnarray}

We analyze the ground-state properties of the system described by the Hamiltonian
(\ref{H}) resorting to the Diffusion Monte Carlo (DMC) technique. The guiding wave
function is of Jastrow type, with a two-body correlation terms corresponding at
short distances to a solution of the two-body scattering problem and smoothly
matched to a long-range hydrodynamic asymptotics at larger distances (further
details in Ref.~\cite{Astrakharchik06}). The parameters of the guiding wave function
are optimized by energy minimization in variational Monte Carlo calculations. The
optimal guiding wave function is then used as an input to the DMC algorithm, which
produces exact (apart from some statistical uncertainty) ground-state energies in a
system of bosons. The pair distribution function and static structure factor are
calculated using the technique of pure estimators \cite{Casulleras95}. The one-body
density matrix is estimated using the extrapolated estimator using DMC and VMC
results \cite{Casulleras95}. As finite-size effects are very important for the case
of a dipolar interaction potential \cite{Astrakharchik06}, we perform calculations
with system sizes $N=19;25;35;50;70;100;200$ and do extrapolation of the energy to
the thermodynamic limit $N\to\infty$.

\begin{figure}
\begin{center}
\includegraphics[width=0.88\columnwidth]{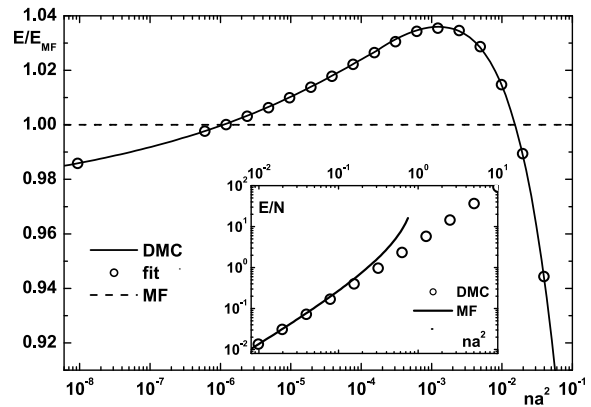}
\caption{Ground state energy per particle as a function of the gas parameter $na^2$.
Inset: in units of $\hbar^2/(Ma^2)$, symbols - DMC data, line - mean-field energy
$E_{MF}$, Eq.~(\ref{Emf}). Main figure: in units of $E_{MF}$, symbols - DMC data,
solid line - best fit, Eq.~(\ref{fit}), dashed line - mean-field energy.}
\label{Fig:E}
\end{center}
\end{figure}

DMC results for the ground-state energy as a function of the gas parameter $na^2$
are reported in Fig.~\ref{Fig:E}. In the inset, we make a direct comparison to the
mean-field energy, Eq.~(\ref{Emf}). We note that while the MF description is in a
qualitative agreement with our data up to large values of the gas parameter
$na^2\lesssim 0.1$, quantitatively there are noticeable differences. In order to
test the quality of analytical equations of state we plot in Fig.~\ref{Fig:E} the
energy in units of the MF energy. Even at the smallest considered density $na^2 =
10^{-8}$ MF theory does not provide a quantitatively correct result. The behavior at
such low densities is expected to be universal and defined by the gas parameter.
Indeed, the energy of dipoles is almost indistinguishable from the energy of both
soft and hard spheres\cite{Pilati05} when $na^2<10^{-6}.$ It is worth noticing that
the situation is different from the 3D case where there is a perfect agreement with
MF for values of the gas parameter $n_{3D}a_{3D}^3\lesssim 10^{-6}$
\cite{Giorgini99}. In a 1D system, the MF result is recovered with an accuracy
better than $10^{-3}$ already for $n_{1D}|a_{1D}|\lesssim 10^{-3}$ \cite{Lieb63}.
The reason for such a difference is that in three- and one- dimensional systems the
expansion comes in terms of the gas parameter, whereas in 2D the expansion is done
in terms of $1/\ln na^2$, and corrections to the energy include slowly converging
terms $\ln |\ln na^2|$. This means that for experimentally interesting densities
$na^2\gg 10^{-8}$, one must consider additional terms in the equation of state
(\ref{Emf}). We tested a number of equations of state present in the literature
\cite{BMF} and none of them is able to reproduce correctly the DMC value of the
energy even at $na^2=10^{-8}$. Thus, the only remaining way to obtain a
quantitatively correct description of a 2D dipolar gas is to perform a microscopic
study. We fit the DMC equation of state in the density range $10^{-6}<na^2<10^3$
using the following function
\begin{eqnarray}
\frac{E}{N}\!=\!
\left\{
\begin{array}{ll}
\frac{\hbar^2}{Ma^2}\exp(a_0\!+\!a_1l\!+\!a_2l^2\!+\!a_3l^3\!+\!a_4l^4), & n>n_c\\
\frac{2\pi\hbar^2n}{M}\left(b_0+b_1l+b_2\ln l+b_3\ln\ln l\right), & n\le n_c
\end{array}
\right.
\label{fit}
\end{eqnarray}
where $l=|\ln na^2|$. The best fit gives $a_0=1.4552(4)$, $a_1=-1.3033(1)$,
$a_2=0.01267(2)$, $a_3=7.7(2)\cdot 10^{-5}$, $a_4=-2.18(4)\cdot 10^{-5}$,
$b_0=-2.225(20)$, $b_1=1$, $b_2=3.294(5)$, $b_3=-6.65(1)$, $n_c=1.8\cdot 10^{-4}$.
Numbers in parentheses indicate the uncertainty of the fitting procedure. The
resulting curve is shown in Fig.~\ref{Fig:E}.

\begin{figure}
\begin{center}
\includegraphics[width=0.88\columnwidth]{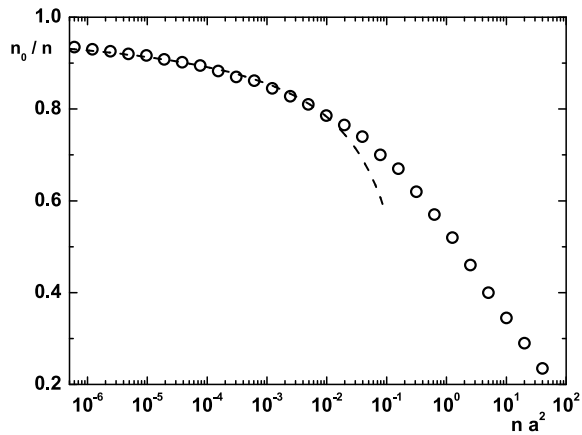}
\caption{Condensate fraction as a function of the gas parameter $na^2$, symbols -
DMC data, dashed line - perturbative low density prediction, Eq.~(\ref{CF}).}
\label{Fig:CF}
\end{center}
\end{figure}

The condensate fraction $n_0/n$ is obtained from the non-diagonal long-range
asymptotics of the one-body density matrix \cite{Astrakharchik06}. In
Fig.~\ref{Fig:CF}, we plot results of $n_0/n$ as a function of the gas parameter. We
find that the predictions of Bogoliubov theory, Eq.~(\ref{CF}), are in a good
agreement in the weakly interacting regime. This should be contrasted with the
energy, where the difference at the same density is significative. Nevertheless, at
larger densities deviations are found. We note that at the densities of physical
interest for excitonic systems $na^2\approx 1-10$, the condensate is highly
suppressed, making the applicability of the Gross-Pitaevskii equation, where the
system is thought to be fully condensed, hardly possible. Moreover, in this regime
it is expected that the universal description in terms of the $s$-wave scattering
length breaks down and the properties of the system depend crucially on details of
the potential.

\begin{figure}
\begin{center}
\includegraphics[width=0.88\columnwidth]{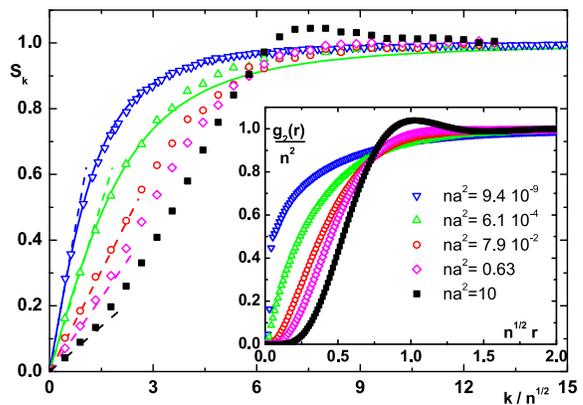}
\caption{Static structure factor: symbols - DMC data,
dashed lines - the low-momenta phononic behavior, Eq.~(\ref{Skph}), solid
lines - Feynman prediction assuming Bogoliubov excitation spectrum
Eq.~(\ref{SkBog}). Inset: pair distribution function. In descending order of the
slope at $r=0$: $na^2 = 9.4\cdot 10^{-9}$; $6.1\cdot 10^{-4}$; $7.9\cdot 10^{-2}$;
$0.63$; $10$}
\label{Fig:Sk}
\end{center}
\end{figure}

In Fig.~\ref{Fig:Sk}, we plot the static structure factor $S_k$ calculated in a
system with $N=200$ particles in a wide range of the gas parameter $9.4\cdot
10^{-9}\le na^2\le 10$. The static structure factor is obtained as a Fourier
transformation of the pair distribution function. The low-momentum part of $S_k$ is
in agreement with the linear phonon prediction (\ref{Skph}), shown in
Fig.~\ref{Fig:Sk} with short-dashed lines. The slope is proportional to the speed of
sound $c$, which, in Fig.~\ref{Fig:Sk}, is calculated from the fit~(\ref{fit}). We
see that for the two smallest considered densities Bogoliubov theory of weakly
interacting Bose gas provides a relatively good description of the static structure
factor. In the inset of Fig.~\ref{Fig:Sk} we show the characteristic behavior of the
pair distribution function $g_2(r)$. In the regime of low densities $g_2(r)$ goes
smoothly from zero (the dipole potential is highly repulsive when $r\to 0$) to the
constant value $g_2(r)\to n^2, n^{1/2}|r|\gg 1$, which is a typical behavior for a
liquid (gas). We note that in this regime at small distances the pair distribution
function is proportional to the square of the two-body scattering solution
$f_{TB}(r)$: $g_2(r)=const\cdot |f_{TB}(r)|^2$, while the coefficient of
proportionality is a many-body property and can not be extracted from the two-body
problem. At large densities $na^2\approx 1$ a peak at mean interparticle distance
$r=n^{-1/2}$ starts to be formed, which shows that the system is more correlated in
this regime. Indeed, it was found\cite{Astrakharchik06,buchler06} that the system of
dipoles experience a quantum phase transition to a crystal if the density is greatly
increased.
%

Differences in the equation of state have consequences also in trapped systems,
leading to different values of the release energy, the size of the cloud, {\it etc}.
One of the most precise techniques, which can be used for testing the equation of
state, is the measurement of the frequency of the ``breathing'' mode produced by a
sudden change of the frequency of the harmonic trapping in the 2D plane. In order to
calculate the frequency of this mode in a trapped system we have used the local
density approximation (LDA) \cite{Astrakharchik06a}. Within this approximation all
properties depend on a characteristic parameter $Na^2/a_{ho}^2$, where $a_{ho}$ is
the oscillator length of the confinement in the 2D plane. The frequencies of the
``breathing'' mode in a 2D system were calculated in Ref.~\cite{Astrakharchik06a}
using scaling approach and assuming that the amplitude of oscillations is small. In
a symmetric trap this frequency is given by
\begin{eqnarray}
\Omega^2 = -2 {\langle r^2\rangle}\left/
{\frac{\partial\langle r^2\rangle}{\partial \omega^2}}\right.
\label{Omega}
\end{eqnarray}
The mean square size of the cloud is calculated within LDA assuming that $\langle
r^2\rangle \gg a_{ho}^2$. This condition is easily satisfied if the number of
dipoles in the trap $N$ is large.

\begin{figure}
\begin{center}
\includegraphics[width=0.88\columnwidth]{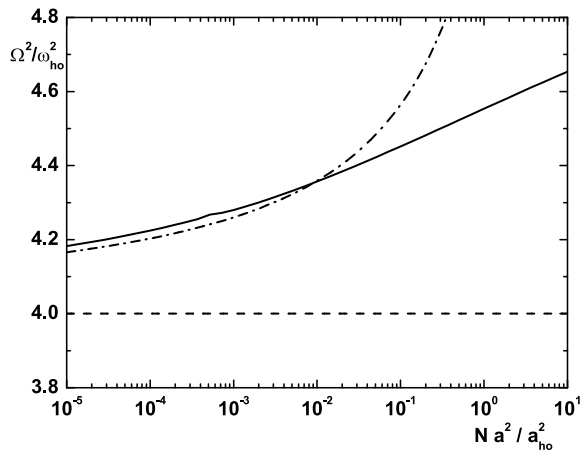}
\caption{Reduced square $\Omega^2/\omega^2_{ho}$ of the lowest monopole frequency
(``breathing'' mode) in an isotropic trap, calculated as Eq.~(\ref{Omega}), for the DMC
(solid line), MF (dash-dotted line) equations of states as a function of LDA
parameter. For comparison we plot $n\to 0$ asymptotic
$\Omega=2\omega_{ho}$ (dashed line).}
\label{Fig:Freq}
\end{center}
\end{figure}

In Fig.~\ref{Fig:Freq}, we show frequencies of the ``breathing'' mode $\Omega$
obtained for different equations of state. In the regime of weak interactions
$Na^2/a_{ho}^2\to 0$ the frequency goes asymptotically to a constant value
$\Omega=2\omega_{ho}$. This value corresponds to the mean-field result $\mu=gn$,
with a coupling constant $g$ independent of density\footnote{The frequency
$\Omega=2\omega_{ho}$ is characteristic for an ideal Fermi gas. In a 2D system
$\mu^{IFG}\propto n$ similarly to low density mean-field limit.}. Differently to
three- and one- dimensional systems, the purely-two-dimensional regime of constant
$\Omega$ can not be probably reached in experiments, since the system should be
extremely dilute. Indeed, even for the smallest considered values of the LDA
parameter $Na^2/a^2_{ho} = 10^{-5}$ the square of the frequency $\Omega^2\approx
4.2\omega_{ho}^2$ is considerably different from the limiting one. Already for this
value of the LDA parameter we find differences between the DMC and mean-field
equations of state. This discrepancy becomes more evident when the system enters
into a regime of stronger interactions $Na^2/a^2_{ho}
\approx 10^{-1}$.
On the other hand, in the characteristic densities of
excitonic experiments $na^2\approx 1-10$ the mean-field description,
Eq.~(\ref{Emf}), fails completely and data should rely on DMC results.
We note that differences of several percent can be resolved in present
high-precision experiments with cold gases (see, for example,
Ref.~\cite{Altmeyer06}).

In conclusion, we have performed a fully microscopic description of a weakly
interacting system of dipoles using the DMC method. The ground-state equation of state,
as well as results for the static structure factor, pair distribution function and
condensate fraction are reported. It has been shown that Bogoliubov theory provides
a good description of correlation functions in the dilute regime. For a harmonically
trapped systems we calculate the frequency of the ``breathing'' mode using local
density approximation. This frequency provides a sensitive tool for testing the
equation of state and can be measured in near future in experiments. An important
conclusion of the present work is the failure of the commonly used Gross-Pitaevskii
mean-field theory (differently from 3D and 1D systems) to describe quantitatively
energetic properties of a weakly interacting 2D system of dipoles. This failure is
present already in the universal regime, {\it i.e.}, where the shape of the
interaction potential is no longer important, suggesting that special care should be
taken in the description of properties of any 2D dilute Bose system.

The work was partially supported by (Spain) Grant No. FIS2005-04181, Generalitat de
Catalunya Grant No. 2005SGR-00779 and RFBR. G.E.A. acknowledges post doctoral
fellowship by MEC (Spain).


\end{document}